\documentstyle[amssymb,multicol,epsf,aps]{revtex}
\def\dbarrm {{\mathchar'26\mkern-11mu{\rm d}}}                         %
\renewcommand{\S}{S_{\rm ep}}
\newcommand{\Tbg}{T_{\rm cb}}

\renewcommand{\TH}{T_H}
\newcommand{\BEQ}{\begin{equation}}
\newcommand{\EEQ}{\end{equation}}
\newcommand{\BEA}{\begin{eqnarray}}
\newcommand{\EEA}{\end{eqnarray}}

\renewcommand{\d}{{\rm d }}

\newcommand{\I}{{S_{BH}}}
\newcommand{\p}{\partial}

\input{psfig}
\begin{document}
\draft
\title{Thermodynamics of black holes: an analogy with glasses}
\author{Th.~M.~Nieuwenhuizen}
\address{Van der Waals-Zeeman Instituut, University of Amsterdam
\\ Valckenierstraat 65, 1018 XE Amsterdam, The Netherlands}
\date{July 12, 1998; e-mail: nieuwenh@phys.uva.nl}
\maketitle
\begin{abstract}
The present equilibrium formulation of thermodynamics for black holes
has several drawbacks, such as assuming the same temperature for
black hole and heat bath.
Recently the author formulated non-equilibrium thermodynamics for
glassy systems.
This approach is applied to black holes, with
the cosmic background temperature being the  bath temperature,
and the Hawking temperature the internal temperature.
Both Hawking evaporation and
 absorption of background radiation are taken into account.
It is argued that black holes did not form in the very early universe.
\end{abstract}
\begin{multicols}{2}

Black holes are singular cosmological objects, with very
strong gravitational forces. Nothing, even light, can escape from it
at sizeable rates.
Many indications point at their presence in the universe.
In the center of our own galaxy there is probably a black hole.

Thermodynamics is the old science that describes the energy balance
of systems, ranging from steam machines to crystals and stars.
It has been a challenge to find out whether an uncommon object as a
black hole is governed by these universal laws.
Attempts in this
direction, started by Bekenstein~\cite{Bekenstein}, will be reviewed.
For introductory texts on the subject, see e.g. ~\cite{Wald}
\cite{Jacobson}\cite{vanDongen}.
A solution will be proposed that is based on the newly formulated
thermodynamics for glasses.

A black hole has ``no hair'', i.e. it can be
 characterized by a few parameters, namely its
mass $M$, charge $q$ and angular momentum $J$.
This is reminiscent of fluids, that can be characterized by
temperature and pressure.
It is known for long that the energy $U=Mc^2$ satisfies
{}~\cite{BCH}
\BEQ\label{1bhd}
\d U= \frac{\kappa}{8\pi} \d A+\Omega\cdot \d J+\phi\d q
\EEQ
where $\kappa$ is the surface tension, $A=4\pi R_s^2$ the area
expressed in the Schwarzschild radius $R_S$, $\Omega$
the horizon's angular velocity, and $\phi$ the
electrostatic potential at the horizon.
This law holds when adding matter to one given black hole,
but also when comparing two different black holes. These two very different
applications suggest a universal validity, and a thermodynamic description.

Classical black holes cannot decrease their surface area~\cite{MTWheeler},
a property reminiscent of the entropy of a closed system.
This analogy motivated Bekenstein ~\cite{Bekenstein}
25 years ago to formulate the laws of  black holes mechanics in a
thermodynamic framework. He introduced as entropy the area in dimensionless
units, so divided by the square of
Planck's length $L_P=\sqrt{\hbar G/c^3}$.
There was still freedom to choose a multiplicative
constant, now known to be $k_B/4$. This leads to the ``information''
entropy
\BEQ \I=\frac{k_B A}{4L_P^2}=\frac{\pi R_S^2k_Bc^3}{\hbar G}\EEQ
The presence of $\hbar$ calls for a quantum mechanical
interpretation.  Not much later Hawking
demonstrated the quantum evaporation of black holes~\cite{Hawking}.
This  underlined the relevance of  Bekenstein's approach.
The black hole radiates as a black body at Hawking temperature
\BEQ
\TH=\frac{\hbar G\kappa}{2\pi c^3k_B}=\frac {\hbar c^3}{8\pi GMk_B}\EEQ
where the second equality holds for a non-rotating, neutral black hole,
having $R_S=2GM/c^2$.
All possible particles are emitted at this temperature; for large
black holes, however, $\TH$ is so small, that in practice
only massless particles  (photons, neutrino's, gravitons) are emitted.

Between these two fundamental steps, Bardeen, Carter and Hawking~\cite{BCH}
had  formulated ``the four laws of black hole dynamics''.
The zeroth law states that the surface tension $\kappa$ is constant
at the black hole surface, just like
the temperature is the same everywhere in an
equilibrium system. The first law is given in eq. (\ref{1bhd}).
Since the last two terms corresponds to
work terms, one may write this relation in the suggestive form
\BEQ \label{1bhth}
\d U=\TH\d \I+\dbarrm W \EEQ
This formulation  is  sometimes referred to as the first law of
 black holes thermodynamics.
Bekenstein had also discussed the {\it generalized second law}
\BEQ\label{gen2law}
 \d\I+\d S_{m}\ge 0 \EEQ
where $S_{m}$ is the entropy of the matter outside the black hole.
The third law states that ``optimal'' black holes,
the ones that have $\kappa=T_H=0$, cannot be reached
by a finite number of steps~\cite{BCH}\cite{Israel}.

{}From the view point of a condensed matter physicist, the
literature on black hole thermodynamics is somewhat confusing.
First of all, one should define the system for which a
thermodynamic description is to be given.
This is rarely done;
 we shall see below what consequences this has.
A natural choice is to consider as system the black hole
and a ``Gedanken'' sphere around it of, say,
a hundred times the Schwarzschild radius.
One could also consider the whole
universe as an isolated container.

Our next objection concerns the formulation of the first law in black
hole literature. The standard formulation is
\BEQ \label{1lth}\d U=\dbarrm Q+\dbarrm W \EEQ
saying that the increase of the system energy equals the heat
added to the system and the work done on the system.
$\dbarrm Q$ has to be determined for the system under consideration.
 The second law only says that
heat must flow from high to low temperatures, which requires that
\BEQ \label{2law}\dbarrm Q\le T\d S \EEQ
The equality sign holds if and only if there is equilibrium.
It is seen that eq.  (\ref{1bhth}) {\it is not the first law
of thermodynamics},
it is obtained from it after identifying $T=\TH$, $S=\I$
and inserting  $\dbarrm Q=T_H\d \I$ from the second law.
This may seem plausible, but it is not: one has assumed equilibrium
of the whole system at temperature $\TH$, which is not present.
It is indeed well known that the ``Bekenstein''
specific heat $C_{\rm Bek}\equiv \d U/\d \TH$ is negative, so even if
equilibrium at $T_H$ were present, fluctuations would drive
the system away from it.
As eq. (\ref{1bhth}) follows from solving dynamical equations,
there is nothing wrong with it, but
{\it it has no foundation within standard thermodynamics}.

In black hole literature it is often stated that the entropy cannot decrease.
Let us recall that eq. (\ref{2law}) only requires that for a closed system.

Having defined the system, one should discuss its
entropy. For the Gedanken sphere with the black hole in it,
eq. (\ref{1lth}) applies. Notice that the entropy of eq. (\ref{2law})
is the one that belongs to the same system, $ S=\I+S_m^{\rm Gs}$.
The latter is the entropy of the cosmic background matter
outside the black hole but inside the Gedanken sphere, and
expected to scale with the sphere's volume.
There is no justification for including the entropy (or the energy)
of matter outside the sphere.
The  radiation generated by the hole will quickly leave the system
and go to the heat bath around it; this is described by a $\dbarrm Q<0$.

If, on the other hand, the whole universe is considered as system,
then $\dbarrm Q=0$. If no work is done, this implies that $\d U=0$,
saying that  energy radiated from the hole is still inside
the system. In that case eq. (\ref{1bhth}) does not describe
the change of the system's energy, it only says something about the
black hole. The total entropy is now $S=\I+S_m$, and
the second law indeed says that $\d S\ge 0$. 

We conclude that eq. (\ref{1bhth}) and (\ref{gen2law}) should not be
applied simultaneously: they refer to different cases.
In practice this means: different time scales. When only the black hole
and its Gedanken sphere are considered, this describes
the radiation emitted in a time $\d t$.
When considering the change in entropy of the whole universe,
one tacitly assumes time scales so
large that the emitted radiation has come in equilibrium.

A final, severe, objection against the current formulation of
thermodynamics for black holes is: {\it  what is the heat bath?}
By considering $T_H$ within thermodynamics, this is by definition
the bath temperature, and normally also the temperature of the object.
This can only apply to a black hole in
equilibrium with its own Hawking radiation, which is an unstable
and thus unphysical situation; it can also not deal with black holes
of different size.
Physically there is one, and only one choice for the bath: for a black hole
that has swallowed all matter around it, the heat bath is the cosmic
background radiation, that presently has temperature $\Tbg=2.73\,K$.
So the actual problem deals with a system of which the dynamics
prefers to ``live'' at a second temperature, namely $\TH$.
This  calls for a two-temperature description.

Recently the author has proposed a thermodynamic description of the
glassy state~\cite{Nthermo}\cite{NEhren}\cite{Nhammer}.
The essential point is that, as there is no equilibrium,
time has to be kept as additional  parameter.
Within thermodynamics a  more useful
extra variable is the effective temperature $T_e(t)$.
Whereas the fast processes
are at equilibrium at the bath temperature $T$, the slow or
configurational modes are at a quasi-equilibrium at $T_e(t)$.
In glasses $T_e(t)$ exceeds $T$. Indeed, in the glass
formation process by cooling from high temperatures, $T_e(t)$
has lost track of $T(t)$, and is since then lagging behind, trying
to reach it in the very remote future. By eliminating $t$ one may specify
the cooling trajectory by a function $T_e(T)$.
By doing smoothly related cooling experiments at a set of pressures
$p_i$ one defines a surface $T_e(T,p)$ in $(T,T_e,p)$-space.
To cover that space, many experiments are
needed, e.g. at different pressures and cooling rates.
Alternatively, one could keep one given system under fixed external
conditions, and consider its aging behavior.
These two options are quite analogous to the
ones for black holes~\cite{BCH},  mentioned below eq. (\ref{1bhd}).

 The fast and slow modes do not  only have
their own temperature, they also have their own entropy. The fast modes have
the {\it entropy of equilibrium processes} $\S$,
while the slow modes involve the
``configurational'' or ``information'' entropy or ``complexity''
${\cal I}$.
The total entropy is $S=\S+{\cal I}$.
The basic point has been the expression for the change in heat
\BEQ \label{glassy2law}
\dbarrm Q=T\d\S+T_e\d{\cal I}
\EEQ
which satisfies (\ref{2law}) since $T_e>T$ and $\d\I<0$.
The latter holds since in the course of time the system will go
to lower, less degenerate modes.
In combination with the  first law this  yields the ``apparent''
specific heat $C\equiv \p U/\p T|_p=T\p\S/\p T+T_e\p{\cal I}/\p T$.
Since both entropies are functions of $T$ and $T_e(T,p)$,
this can be written as $C=C_1+C_2\, \p T_e/\p T$,
a form postulated by Tool ~\cite{Tool} and often used
to describe the behavior in the glass formation region.
Since $T_e$ is a decreasing function of time, $\p T_e/\p T=\dot T_e/\dot T$
is positive in cooling, but negative for subsequent
heating in the glassy state. Only when reaching the liquid state
again, it becomes positive and actually exhibits an overshoot.
In simple models $C_1$
vanishes in the glassy regime, so $C$ is negative upon heating.
In realistic glasses $C$ is larger in cooling than in heating,
which is the same effect on top of a background $C_1$, that
arises from uninteresting, fast equilibrium processes.

When applying these ideas to black holes, the bath is the
universe filled with cosmic background radiation, presently having
temperature $\Tbg\approx 2.73\,K$.
The system's internal, effective temperature is
the Hawking temperature. This is in agreement
with the time scale argument. Black holes heavier than
$10^{-18}M_\odot=10^{15}\,g$
need more time to evaporate than the present age of the universe.
For them the evaporation process, as seen by far-away observers,
is so slow,  that equilibration of the cosmic background radiation
is a fast process.

The slow evaporation processes occur at the Hawking temperature
and have as associated entropy the Bekenstein-Hawking
black hole entropy $\I$, so eq. (\ref{glassy2law}) becomes in this
context
\BEQ \label{dQBH}
\dbarrm Q=T_{cb}\d S_m^{\rm Gs}+T_H\d \I
\EEQ
Because $S_{BH}$ is so large, the entropy of the background
radiation outside the back hole but inside the
Gedanken sphere is negligible, $S_m^{\rm Gs}\ll \I$, implying
$\dbarrm Q=T_H \d \I$. 
Together with eq. (\ref{1lth}) this reproduces (\ref{1bhth}),
but now it has received its non-equilibrium interpretation.
Using eq. (\ref{dQBH}) and $S=S_m^{\rm Gs}+\I$
 the second law (\ref{2law}) implies
\BEQ\label{2lawbh}
(\Tbg-T_H)\d\I \ge 0
\EEQ
Hawking radiation leads to $\d \I<0$. Eq. (\ref{2lawbh}) is thus
satisfied as long as $T_H>\Tbg$, but not below that. One might
think that $\Tbg$ plays no physical role whatsoever, and only shows up
as determinator in the second law.
However, the real point is that we not yet
considered absorption of background radiation by the black hole.
The  absorption rate will be proportional to the
area times the energy density, i.e., $\sim M^2\Tbg^4$.  One might be tempted to
find a time-dependent solution of the Einstein equations
for obtaining the prefactor $\alpha_{abs}(T)$.
However, what is needed is the quantum absorption process. We can solve that
without doing any calculation, because it is the time-reversed evaporation
process.   For non-rotating,
neutral holes Hawking radiation  leads to a mass loss
\BEQ
\dot M=-\alpha_{em}\frac{\hbar c^4}{G^2M^2}
\to \dot T_H=\frac{(8\pi)^3 \alpha_{em}G k_B^3}{\hbar^2 c^5}\TH^4
\EEQ
The dimensionless constant $\alpha_{em}$ depends on the type of particles
present, and their absorption probabilities, called
``oscillator strengths'' in solid state systems.
$\TH$ enters through the Bose-Einstein occupation numbers (for bosons,
in particular photons) or Fermi-Dirac occupation numbers (for fermions).
For an uncharged, non-rotating black hole Page finds
$\alpha=5.246\times 10^{-4}$ in the high-frequency limit, and
$0.181\times 10^{-4}$ in the low frequency limit~\cite{Page}.
For absorption by the black hole of a photon (or a particle) from the
cosmic background, the time-reversed problem shows up.
It thus holds that $\alpha_{abs}(T)=\alpha_{em}(T)$,
no matter the character of the particle content;
for simplicity we shall now replace both by a constant.
The only difference between the
two situations is the temperature occurring in the occupation numbers:
for Hawking emission it is $T_H$,
while for cosmic background absorption it is $\Tbg$.
The combined processes of Hawking emission and background photon
absorption thus yields for a neutral, non-rotation black hole ~\cite{Zurek}
\BEQ\label{balance}
\dot T_H=\frac{ (8\pi)^3 \alpha G k_B^3 }{\hbar^2c^5}(\TH^4-\Tbg^4)
\EEQ
It exhibits an instability at $\TH=\Tbg$,
related to the fact that $C_{\rm Bek}<0$.
If there is equilibrium, and $\Tbg$ is changed a little,
then $T_H$ branches away from it.

There are two regimes. In the ``classical''
regime $\TH<\Tbg$ the black hole absorbs more radiation than is emits.
Its entropy will increase, and $\dbarrm Q=T_H\d \I>0$,
but this is still in accord with the second law (\ref{2lawbh}).
In the ``quantum'' regime $T_H>\Tbg$ the black hole emits
more than it absorbs. Now it holds that $\d \I<0$, confirming
again that heat flows from high to low temperature.

In analogy with glasses, one can define the {\it apparent specific heat}
$C=\p U/\p \Tbg =\dot U/\dot \Tbg$.
For black holes this object is less natural because
the background temperature cannot be changed by hand.
However, $C$ does have a meaning in
our expanding universe. Due to the decrease of $\Tbg$, there will be
less and less background energy to be absorbed. A black hole
will reach its maximal size  at the moment $t=t_0$ where
the temperatures match, $\TH=\Tbg=T_0$; from then on it will shrink.
Around $t_0$ the apparent specific heat
takes a form independent of $\dot\Tbg$, viz.
$C=k_B(t-t_0)/\tau$, with characteristic time scale
$\tau=\hbar/[(16\pi)^2\alpha k_B T_0]$.
In the classical regime ($t<t_0$) $C$ is negative, while
in the quantum regime it is positive.

The third law of thermodynamics
concerns the entropy for $\Tbg\to 0$. We have seen already that
finally all black holes evaporate,  thereby
lowering their configurational entropy very much, in accord
with Planck's third law. What happens ultimately with the black hole
has been the focus of studies by 't Hooft ~\cite{tHooft}.

The entropy change of the universe is found
as for  black body radiation~\cite{Zurek}
\BEQ\label{dSdIGs}
\frac{\d S_m}{\d\I}=\frac{\TH\d S_m}{\d U}=-\frac{\TH\d S_m}{\d U_m}
=-\frac{4\TH(\TH^3-\Tbg^3)}
{3(\TH^4-\Tbg^4)}\EEQ
yielding the entropy production $\dot S=\dot S_m+\dot \I$
\BEQ\label{SprodGs}
\dot S=\frac {\alpha k_B^2}{24\pi\hbar}\,
\frac{(\TH^2+2\TH\Tbg+3\Tbg^2)(\TH-\Tbg)^2}{\TH^3}
\EEQ

Our study of models for glasses has put forward a possible
 universality for fluctuations that arise from
mechanically coupled degrees of freedom. In terms of the four
vectors $M_a=(\Omega_a,\phi)$, and $H_a=(J_a,q)$ we expect
that when one writes $M_a(t,H)=M_a(\TH(\Tbg,H);H)$,
the following relations hold
{}~\cite{Nhammer}
\BEA\label{flucts=}
\chi_{ab}&\equiv&
\frac{\p M_a}{\partial H_b}\Bigl|_{\Tbg}\Bigr.
= \chi_{ab}^{\rm fluct}+\chi_{ab}^{\rm conf}
\\ \label{chifluct}
\chi_{ab}^{\rm fluct}&=&
\frac{\langle \delta M_a(t)\delta M_b(t)\rangle}{\TH(t)};\quad
\chi^{\rm conf}_{ab}=
\frac{\partial M_a}{\partial T_H} 
\frac{\p T_H}{\partial H_b}       
\EEA
The fluctuation term is the quasi-equilibrium expression that could have
been guessed naively, and the configurational term is intuitively also clear.
As there is no equilibrium, we do not expect that the correlation function
$C_{ab}(t,t')=\langle\delta M_a(t)\delta M_b(t')\rangle$
and the response function $G_{ab}(t,t')=\p M_a(t)/\p H_b(t')$
depend solely on $t-t'$. Nevertheless, we do expect the validity
of the quasi-equilibrium fluctuation-dissipation relation
\BEQ
\frac{\p C_{ab}(t,t')}{\p t'}=\TH(t')G_{ab}(t,t')
\EEQ
However, for the specific heat no universal quasi-equilibrium
fluctuation expressions are found in glasses, and we have no reason to
expect them for black holes. This is reassuring in regard of the negative
``Bekenstein'' specific heat. It is a challenge to test these ideas.

Let us now consider the whole universe as our system, so
the entropy of the universe $S_m$ has to be taken into account.
The total entropy is $S=S_m+\I$, while eq. (\ref{glassy2law})
becomes $\dbarrm Q=\Tbg\d S_m+\TH\d\I$. As $\dbarrm Q=0$,
the second law (\ref{2law}) again leads to
(\ref{2lawbh}), but the entropy production
\BEQ\label{Sdotuniv}
\dot S=\frac {\alpha k_B^2}{8\pi\hbar}\,
\frac{(\TH^4-\Tbg^4)(\TH-\Tbg)}{\TH^3\Tbg}
\EEQ
exceeds eq. (\ref{SprodGs}).
The difference is due to equilibration of the emitted radiation
in the universe.
For small black holes, having life time less than the age of the
universe, eq. (\ref{Sdotuniv}) does not apply.
They are fully evaporated before the emitted radiation can equilibrate.

In the early universe $\Tbg$ was large, and it may not have dropped
from the energy balance of the black hole.
Let us estimate the temperature at which the entropy of
ordinary matter and the entropy of the same matter as a black hole
had equal thermodynamic impact
\BEQ \Tbg^\ast S_{\rm star}=\TH\I\to \Tbg^\ast \frac{M}{M_\odot} 10^{58}
=\TH\,  \frac{M^2}{M_\odot^2}10^{77} \EEQ
Using $T_H=T_H^{\odot}M_\odot/M$ we see that the masses drop from
the equality, and we get $\Tbg^\ast = 10^{12}K$,
or an energy of $75\,MeV$. This rough estimate might basically
connect the entropy gain for black hole formation with
disappearance of spontaneous quark-antiquark pair creation
in the early universe.

In conclusion, we have shown that the non-equilibrium
thermodynamics  formulated for glasses
also applies to black holes. It starts by considering the cosmic
background radiation as heat bath, and the Hawking temperature as
an internal temperature of the black hole. It is important to take
into account not only the quantum evaporation of the black hole,
but also its absorption of cosmic background radiation.
Black holes with $T_H>2.73\,K$ evaporate, while
the ones having $T_H<2.73\,K$ (and mass larger than
$2.2\,10^{-8}M_\odot$) absorb more radiation than they emit,
and continue to grow until $\Tbg$ passes through $\TH$.

Our approach incorporates the known properties of dynamics,
and shows how the generalized second law comes into the play.
Both the formation and evaporation of
black holes leads to an increase of the entropy of the whole universe.
Our picture involves the standard zero-entropy formulation of the third law
of thermodynamics, thus putting aside the third law of black hole mechanics.
To the best of our knowledge, there is no contradiction
with the occurrence of negative specific heats.

Let us stress that our approach does not involve
a partition sum, but
merely considers known aspects of the dynamics from a
thermodynamic view point. This is quite reassuring, since outside
equilibrium use of the partition sum would be ill based,
and it would also be ill defined.

An intriguing question is the physical meaning of the black hole entropy.
If we push the analogy with solid state physics further,
we may expect it to be the
logarithm of the number of available states of the matter present in the
black hole. Though the species-part of the entropy is much smaller
than the gravitational part, we see no compelling
reason why the black hole should have ``forgotten''
which particles it has been made of.

\acknowledgments
The author is grateful for discussions with
J.A.E.F. van Dongen, G. 't Hooft, L.J. van den Horn, J.M. Luck,
and H. Verlinde, and for hospitality at CEA-Saclay (France),
where part of this work was done.

\end{multicols}
\end{document}